# Terahertz dynamics of a topologically protected state: quantum Hall effect plateaus near cyclotron resonance in a GaAs/AlGaAs heterojunction

A.V. Stier[1], C.T. Ellis[1], H. Zhang[1], D. Eason[1], G. Strasser[1], B.D. McCombe[1]

T. Morimoto[2], H. Aoki[2] and J. Cerne[1]

[1]Department of Physics, University at Buffalo, The State University of New York, Buffalo, NY 14260, USA

[2]Department of Physics, University of Tokyo, Hongo, Tokyo 113-0033, Japan

We measure the Hall conductivity of a two-dimensional electron gas formed at a GaAs/AlGaAs heterojunction in the terahertz regime close to the cyclotron resonance frequency by employing a highly sensitive Faraday rotation method coupled with electrical gating of the sample to change the electron density. We observe clear plateau- and step-like features in the Faraday rotation angle vs. electron density and magnetic field (Landau-level filling factor), which are the high frequency manifestation of quantum Hall plateaus – a signature of topologically protected edge states. The results are compared to a recent dynamical scaling theory.

The DC quantum Hall effect (QHE) [1] has critically shaped our current understanding of two-dimensional electron gases (2DEGs). This remarkable effect is now known to be the first example of a topologically protected state [2]; the transverse Hall conductivity $\sigma_{xy}$ is quantized due to the suppression of backscattering in the quantum Hall edge channels. In spite of great progress in understanding the physics of the QHE, there are remaining questions, with a particularly important one being how the static Hall conductivity evolves into the dynamical (optical) Hall conductivity, $\sigma_{xy}(\omega)$. This is currently being addressed by experimental [3-7] and theoretical [8, 9] investigations. According to the localization picture, the QHE emerges from the coexistence of localized and delocalized states in disorder-broadened Landau levels [7], with QHE plateau-to-plateau transitions from one quantized value of $\sigma_{xy}$ to another occurring each time the Fermi energy passes from one localized regime of a particular Landau level (LL) to another via delocalized states [10]. At zero frequency (DC), the distinction between localized and delocalized states is clear; other than the topologically protected edge states, only states with localization length $\xi$ larger than the sample size *L* can carry a DC current. For ac driving electric fields, the distinction is less clear as both localized and delocalized states can contribute to the ac conductivity with the localized states oscillating about their localization centers. At sufficiently high driving frequencies, the amplitude of the oscillations of delocalized states becomes smaller than the DC localization length $\xi$, at which point the localized and delocalized states become indistinguishable, and the signature of the QHE, the plateaus, disappears. An analogous result occurs as the sample size is reduced below $\xi$, so the QHE scales with frequency and sample size in similar ways.

Recently, Morimoto et. al. have provided theoretical evidence that the step-like structure of the static Hall conductivity should survive even in the THz regime, despite the fact that the dynamical response is dominated by optical transitions between adjacent LLs rather than ac transport within a LL (see Fig. 1(a)) [8]. A subsequent theoretical work [9] predicts that scaling analysis can be used to understand the frequency dependence of the QHE plateaus into the THz range. Experimental work to test these predictions has been limited. Microwave measurements at magnetic fields where the radiation frequency is far below the cyclotron resonance (CR) frequency ($\omega_c$), have shown plateaus in $\sigma_{xy}(\omega)$ up to 33 GHz [3, 11], and a universal scaling in $\sigma_{xx}(\omega)$ up to 55 GHz [4, 5]. Recently, time domain THz spectroscopy has been used [6] to measure $\sigma_{xy}(\omega)$ in the 0.7-1.2 THz range on an ungated GaAs/AlGaAs heterojunction. That work shows an inflection point in $\sigma_{xy}(\omega)$ around filling factor 2 derived from the deviation from the Drude behavior of fits to the $\sigma_{xy}(\omega)$ data in the region $\omega \approx 0.1\,\omega_c$, well below the CR. However, plateau-like behavior clearly associated with the QHE near CR has yet to be observed and explored in spite of literally hundreds of publications on cyclotron resonance in GaAs 2DEGs since 1985 [12].

We use a highly sensitive polarization-modulation technique to measure the Faraday angle (directly proportional to the Hall conductivity) of a 2DEG at the interface of a GaAs/AlGaAs heterostructure in the THz regime close to $\omega_c$. In the thin film limit, the Faraday rotation, $\theta_F$, is directly proportional to the optical Hall conductivity through $\theta_F(\omega) + i\eta(\omega) \simeq \frac{1}{(1+n_s)c\epsilon_0}\sigma_{xy}(\omega)$. The cyclotron resonance line shape is given from the Drude expression of the optical Hall conductivity $\sigma_{xy}(\omega) = \frac{n\,e^2}{m_{eff}}\frac{\omega_c}{(\omega+i/\tau)^2-\omega_c^2}$. In this region of strong CR absorption and dispersion we have observed clear and robust plateau- and step-like features in the Faraday rotation angle vs. electron density and magnetic field (LL filling factor). These features appear near the expected regions of density and magnetic field that correspond to quantum Hall plateaus at integer filling factors. We attribute these features to the robustness of the topologically protected surface states, which are preserved even in the presence of inter-LL absorption due to suppression of inter-LL backscattering (see Supplementary Information).

We use the combination of monochromatic emission lines from an optically pumped molecular gas laser and a polarization modulation technique [13, 14] (see Supplementary Information for details) to probe $\sigma_{xy}(\omega)$ at two frequencies, 2.52 and 3.14 THz, by measuring the Faraday rotation angle with a sensitivity better than 0.1°. The high signal-to-noise ratio of these measurements has allowed us to investigate the scaling of the $\sigma_{xy}$ plateaus with both frequency and temperature, and to compare it with theory [5, 8, 9].

In a typical QHE system disorder broadens each LL, with delocalized states at the unperturbed energy of each LL and localized states lying on either side [10]. As the Fermi energy is moved through an energy region of localized states, $\sigma_{xy}$ remains constant and the longitudinal conductivity $\sigma_{xx}$ is close to zero, resulting in the well-known QHE plateaus. When a critical energy $E_c$, which is located at the center of each disorder-broadened LL for infinitely large samples, is approached, the localization length $\xi$ diverges as a power law,

$$\xi(E) \sim |E - E_c|^{-\gamma} \quad (eq. 1)$$

where $\gamma$ is the localization critical exponent [5]. In a QHE system, $\xi$ is the typical spatial extent of the wave function at energy $E$. In an infinitely large sample at T = 0 K, the transition from a localized regime for $E < E_c$ (or $E > E_c$) to a delocalized one at $E = E_c$ is sharp. For a finite-size sample with spatial dimension $L$ the states that have $\xi(E) > L$ are effectively delocalized, so that there is a non-zero energy range for the delocalized regime. This implies that the energy width $W$ of the transition from localized to delocalized states is given by $W \sim L^{-1/\gamma}$. Non-zero temperature and finite probe frequencies can also effectively delocalize states; they introduce time scales $\tau_T \sim \hbar/k_B T$ and $\tau_\omega \sim 1/\omega$ respectively. These time scales can be compared to the time scale determined by the localization length $\tau_\xi \sim \xi^z$ with a dynamical critical exponent $z$, to determine their relative importance. Taken together they define an effective system size $L_{\text{eff}} \sim \tau^{1/z}$ that can alter the conductivity of the sample. In general, the components of the conductivity tensor are thought to evolve from the DC to the ac regime as functions exhibiting a scaling behavior. In the DC, zero temperature case the dependence of $\sigma_{ab}$ on energy is $\sigma_{ab}(E) = G\left(\frac{L}{\xi}\right) = G_L(L^{1/\gamma} \delta E)$, where $\delta E = E - E_c$. By introducing frequency and temperature, we can express the scaling functions for the conductivity as

$$\sigma_{ij}(\omega, T = 0) = G_\omega(\omega^{-\kappa} \delta E) \quad (\text{eq. 2})$$
$$\sigma_{ab}(\omega = 0, T) = G_T(T^{-\kappa} \delta E),$$

with a scaling exponent $\kappa = 1/z\gamma$. The scaling behavior of the transition width then is given by $W \sim T^\kappa$ and $W \sim \omega^\kappa$. Non-interacting electron systems [15, 16] as well as systems with short range interactions [17] are known to be governed by a dynamical exponent $z = 2$; long-range interactions change the exponent to $z = 1$ [18]. Studies of the GHz longitudinal conductivity $\sigma_{xx}(\omega)$ reveal a scaling behavior with a critical exponent of $\kappa = 0.5 \pm 0.1$, yielding a dynamical exponent of $z = 0.9 \pm 0.1$ [3, 5]. Although this scaling picture for the localization physics is only valid in the small frequency region ($\omega \ll \omega_c$), it is an intriguing issue to investigate how the localization due to disorder affects the dynamical response around CR.

The main objective of this experimental work is to explore $\sigma_{xy}(\omega)$ of a 2DEG in the THz spectral region close to CR, and thereby to determine the residual effects of the QHE at these high frequencies in the vicinity of electron CR. We present direct measurements that show multiple quantum Hall plateaus in $\sigma_{xy}(\omega)$ near CR. Magnetic field sweeps at constant gate voltage and gate voltage sweeps, which vary the filling factor at constant magnetic field, reveal clear plateaus in the Faraday angle for several integer filling factors. The system studied is a 2DEG formed at the interface of a GaAs/AlGaAs heterojunction ($n_{2D}(V_g = 0) = 5.7 \times 10^{11} \text{cm}^{-2}, \mu = 1.7 \times 10^5 \text{cm}^2/\text{Vs}$ at $T = 77$K. The rotation of the plane of polarization of the electric field in the transmitted light is the Faraday angle, $\theta_F$, which we measured with a combination of a liquid-He cooled Si bolometer and phase sensitive lock-in detection (details see Supplementary Information)[14].

Figure 1 summarizes our key results. The Faraday angle $\theta_F$ measured at 2.52 THz is shown in Fig. 1(b) as a function of magnetic field for a series of constant top-gate voltages. A strong resonance feature, consistent with predictions from the Drude model for CR in a 2DEG, and a smaller feature near 6.2 T, which we associate with residual bulk carriers in the sample substrate, are observed. The inflection point of the large resonance feature marks the position of CR. The shift in the position of the resonance feature to higher magnetic fields with increasing electron density is due to band nonparabolicity [19]. We use this shift of CR (see inset in Fig. 1(b)) to determine $n_{2D}(V_g)$ (details see Supplemental information). This allows us to translate our data into the parameter space $\left(\frac{\omega}{\omega_c}, \nu\right)$, where the cyclotron frequency $\omega_c = \frac{eB}{m_{\text{eff}}}$ and the filling factor $\nu = \frac{n_{2D}h}{eB}$. Figure 1(c) shows $\theta_F$ measured at $\omega_{THz} = 3.14$ THz, translated into the parameter space $\left(\frac{\omega}{\omega_c}, \nu\right)$. The orange arrow in Fig. 1 (b)-(d) marks the most prominent feature near $\nu = 10$, which can be readily observed in the raw data at 2.52 and 3.14 THz. Theoretical calculations [8] based on the sample parameters are shown in Fig. 1(d) and are in good qualitative and quantitative agreement with the measurements. Specifically, simulation and experiment agree in the magnitude of $\theta_F$ as well as in the shape of the ridge line along the filling factor axis at a particular value of $\frac{\omega}{\omega_c}$. In Fig. 2(a) we compare $\theta_F$ measured for magnetic fields slightly above CR $\left(\frac{\omega}{\omega_c} = 0.985\right)$ at two different excitation frequencies. We observe very pronounced cusps in $\theta_F$ slightly above $\nu = 10$, behavior clearly outside that of the semi-classical Drude model. Less pronounced features in this wide stepped data mesh can be seen for the higher frequency in the vicinity of $\nu = 6$ and $8$. These cusps are predicted by theory (see mesh lines in Fig. 1 (d)). As $\theta_F$ is directly proportional to $\text{Re}(\sigma_{xy})$, structures in $\theta_F$ imply structures in $\sigma_{xy}(\omega)$. The exact position and shape along the filling factor axis is determined by the evolution of the DC quantum Hall effect plateau into the AC regime. The appearance of a plateau-like feature at a non-integer $\nu$ (see also features in Fig. 2 (b)) is consistent with earlier results [6], which have been explained as an effect of the interplay of $\xi(E)$ with a frequency-dependent length $L_\omega \sim \omega^{-1/z}$, which is the distance an electron travels over one cycle of the driving ac field [9]. Note that the plateau value is not quantized in the ac regime and the plateau-like structure is in general not perfectly horizontal, in contrast to the DC QHE. Interestingly, the cusps for the higher frequency are more pronounced and the overall $\theta_F$ is larger. This may seem counterintuitive since one expects the QHE to diminish at higher frequencies. The plateau features close to CR are dominated by the broadening $\Gamma$ of the LL density of states normalized by the cyclotron energy $\hbar\omega_c$ [8, 20]. For increased cyclotron frequency, $\frac{\Gamma}{\hbar\omega_c} \sim \frac{1}{\sqrt{B}}$ decreases and therefore produces a more distinct plateau. The inset for Fig. 2(a), shows calculated $\theta_F$ in the vicinity of an integer filling factor $\nu$ close to CR for two effective scattering strengths that match the experimental parameters for the two chosen laser frequencies. The calculation quantitatively agrees with our experimentally determined $\theta_F$, and it shows a qualitative change in shape of the plateau that is consistent with the above picture.

We have further investigated $\theta_F$ as a function of top gate voltage at a series of constant magnetic fields above CR. As in the DC QHE, this measurement corresponds to changing $\nu$ at constant $B$. Before we discuss the results in detail, we compare DC with AC data in Fig. 2(c). Similar to results from [6], we find that AC and DC results track each other largely. However, the small number of data points in the experiments for swept B and constant $V_g$ allow only the observation of few deflection points from a straight line. Detailed study of $\theta_F\left(\frac{\omega}{\omega_c},\nu\right)$ is possible for the case of fixed B fields and swept $V_g$. Figure 2(b) shows the Faraday rotation as a function of $\nu$ at various fixed magnetic fields. We find clear plateaus for $\nu = 6$ and 12 and less pronounced plateaus for $\nu = 8$ and 10. We believe that the origin of this behavior comes from the fact that in our heterostructure, two subbands are occupied for the carrier densities under consideration. We elaborate on the details of this behavior in the Supplemental information. With increasing magnetic field (decreasing $\frac{\omega}{\omega_c}$), the magnitude of $\theta_F$ follows the Drude resonance background, and the overall magnitude of $\theta_F$ decreases at a given $\nu$ as $\omega_c$ moves away from the probing frequency. We observe a pronounced plateau feature for $\nu = 6$ for all B fields. The overall magnitude of the feature diminishes for decreasing $\frac{\omega}{\omega_c}$, which is directly correlated to the Drude factor in the expression for $\theta_F(\omega)$. The size of the features that originate from the quantum Hall effect scale with the Drude factor and therefore can be studied easier close to CR. The plateau for $\nu = 8$ is only weakly established as a slight inflection in the $\theta_F(\omega)$ vs. $\nu$ data. The next higher filling factor plateau $\nu = 10$ is clearly established close to CR and then quickly smoothens out to be similarly pronounced as the $\nu = 8$ plateau for higher B fields. The plateau for $\nu = 12$ appears gradually for magnetic fields above ~6.5 T $\left(\frac{\omega}{\omega_c} \sim 0.99\right)$, which contrasts with the pronounced plateau-like structure for $\nu = 6$ that is present for all values of $\frac{\omega}{\omega_c}$ studied. Very close to CR, heating effects of the 2DEG through the absorption of THz radiation are likely the dominant mechanism responsible for the disappearance of the plateaus. As we show in figure 2(d), the plateau width and the plateau position depend on $\left(\frac{\omega}{\omega_c}\right)$ (see SI for detailed description of determination). Our current understanding is that the position of the plateau feature is based on the interplay of localization length and effective sample size. Morimoto et. al [9] predict a gradual shift of the plateau to higher $\nu$ with increasing $\left(\frac{\omega}{\omega_c}\right)$. This is observed for the $\nu = 6$ plateau for $\left(\frac{\omega}{\omega_c}\right) <$ 0.93. However, the shift of the plateaus towards lower filling factor for $\left(\frac{\omega}{\omega_c}\right) > 0.93$ and in particular the rapid shift of the $v = 6$ plateau position towards lower filling fraction close to CR is not understood. We emphasize that this observations is not due to the change in effective electron mass with filling factor due to band non-parabolicity in semiconductor 2DEGs [21]. We do not expect the effective sample size to change rapidly with $(\frac{\omega}{\omega_c})$, since $L_\omega$ is a smooth function of $\omega$. Therefore, we attribute the non-monotonic shift of the plateau to a non-monotonic change in the localization length very close to CR, or to an effect of the strong CR absorption, which modifies the scaling behavior. In addition, although the shifts of the positions of the plateau structures are qualitatively similar, the widths behave differently. The plateau widths exhibit a weak dependence on $(\frac{\omega}{\omega_c})$, while the plateau width for $v = 12$ rapidly decreases with increasing $(\frac{\omega}{\omega_c})$, going to zero at $\left(\frac{\omega}{\omega_c}\right) = 1$ as compared to the width of the $v = 6$, which only slightly decreases towards $\left(\frac{\omega}{\omega_c}\right) = 1$.

The observed plateau structures are surprisingly sensitive to temperature. Figure 3 shows the $v = 6$ plateau at $\left(\frac{\omega}{\omega_c}\right) = 0.97$ at three different temperatures. We observe a slight shift of the plateau-like feature to lower $v$ with increasing temperature, as well as a broadening and decrease in both the height and width of the step. The width of the plateau-like region is plotted in the inset. We can describe the diminishing of the plateau width using a power law with an exponent of $\kappa = 0.5 \pm 0.1$, which is consistent with temperature scaling behavior in the GHz regime [5] and points towards an electron-electron interaction mechanism.

In conclusion, we have measured the THz Faraday rotation in a GaAs 2DEG near the magnetic field/frequency region of cyclotron resonance. In addition to the expected strong response, with a dispersive line shape, due to the resonant chiral optical transitions between adjacent Landau levels that can be described by a Drude model, we have observed clear step-like and plateau-like features close to integer filling factors, which are analyzed and compared to recent calculations of the THz $\sigma_{xy}(\omega)$ for a 2DEG. There is a qualitative difference in the underlying physics in the frequency regime of our measurements and previous GHz and THz studies of $\sigma_{xy}(\omega)$; in those experiments the only allowed optical transitions correspond to *intra*-LL processes, whereas strong *inter*-LL transitions (CR) dominate the present experiments ($\omega \sim \omega_c$). Our measurements reveal a rich structure corresponding to the dynamical integer QHE in a parameter space that has not been probed before. While various observed features have been predicted by, and agree qualitatively with, theory [8, 9], some observations require further theoretical work to achieve a full understanding. The temperature scaling behavior for step-like features close to even filling factor in the CR regime, and the estimated scaling exponents are consistent with theoretical predictions based on electron-electron interactions.

This work was supported by NSF-DMR1006078 (CTE & JC), NSF-MWN1008138 (AVS & BDM), HA has been supported in part by a Grant-in-Aid for Scientific Research No. 23340112 from MEXT and TM by JSPS.

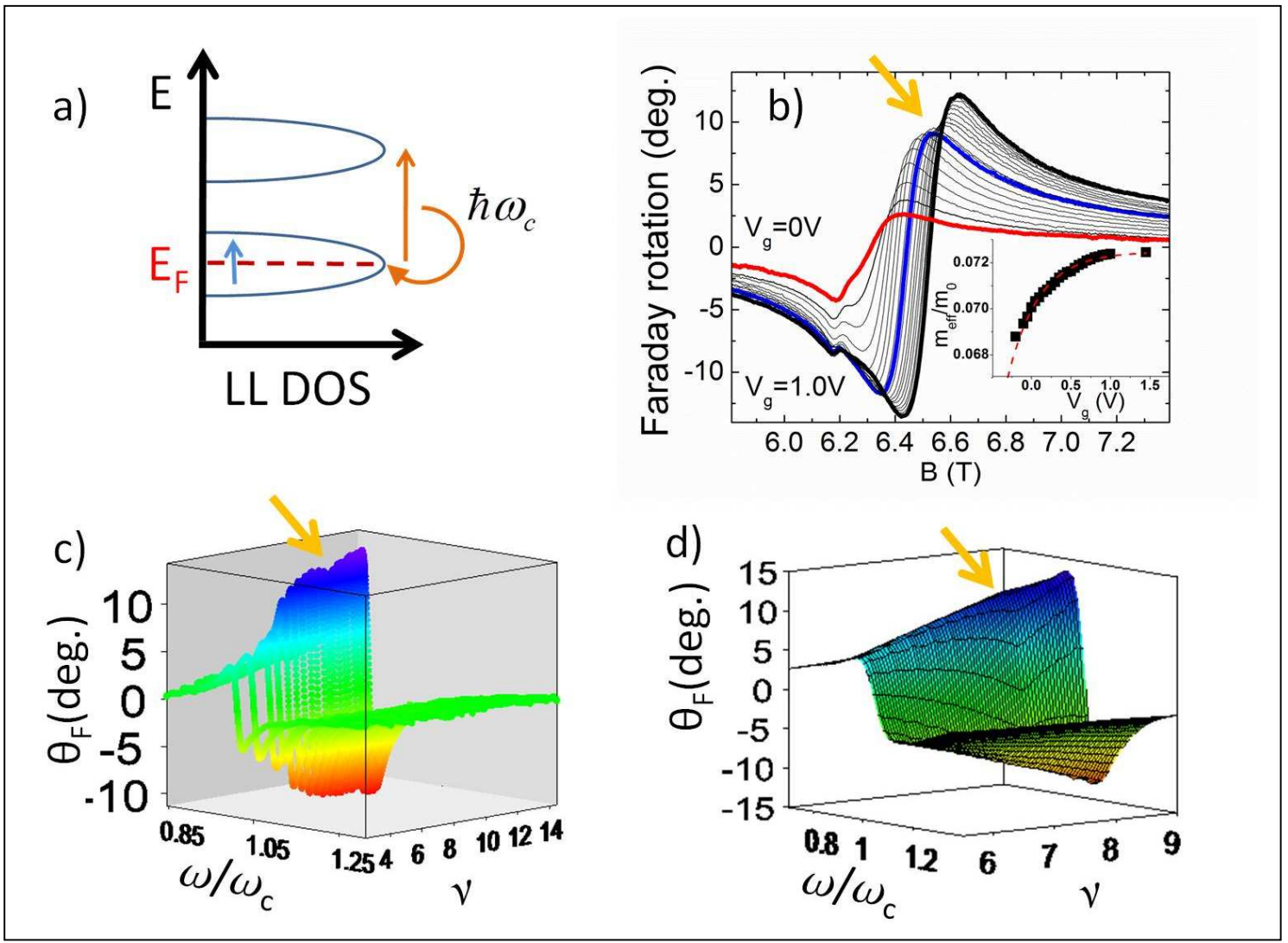

Figure 1.

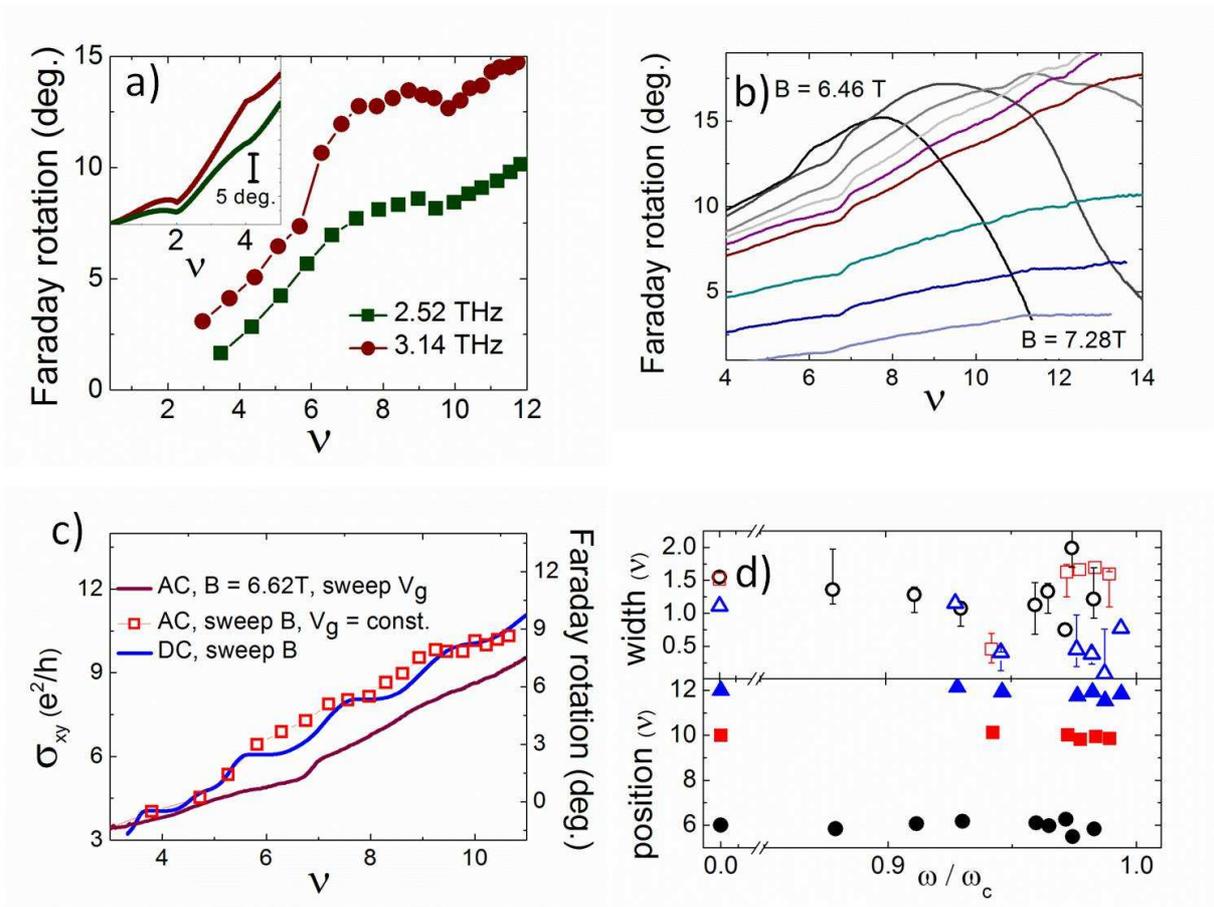

**Figure 2.**

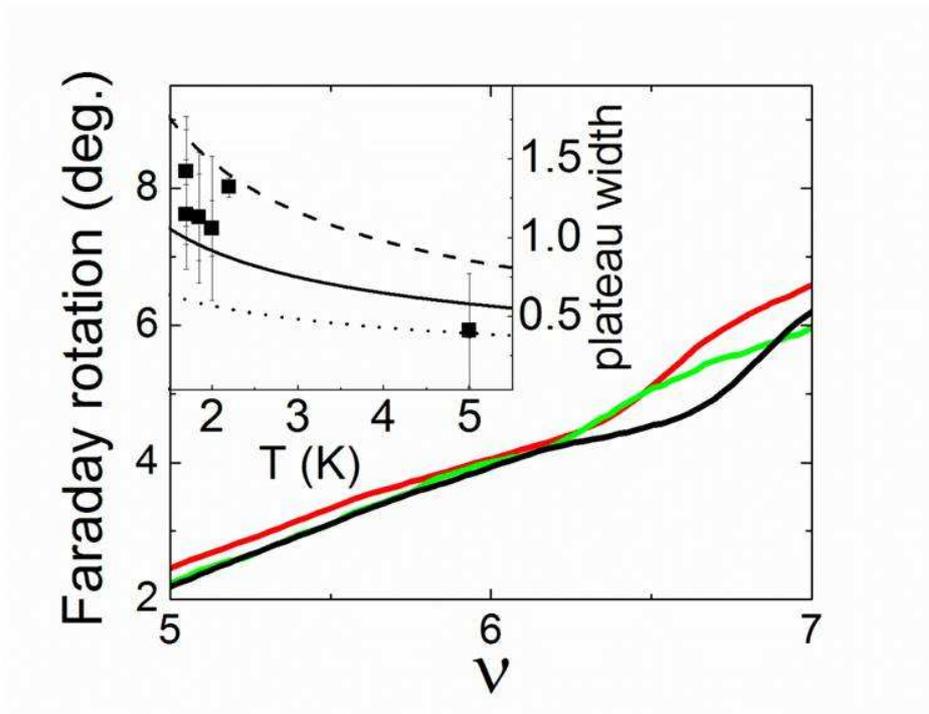

**Figure 3.**

**Figure captions**

**Fig.1:** a) Cartoon of Drude-like intra-band transitions (blue arrow) and circularly active CR inter-band transitions (orange arrow) between adjacent LL. b) Faraday rotation for $\omega_{THz} = 2.52$ THz versus magnetic field B at various, fixed gate voltages $V_g$. The large background is due to CR. Inset shows shift of effective electron mass as function of $V_g$. c) Faraday rotation for $\omega_{THz} = 3.14$ THz versus B at various fixed $V_g$ translated into the parameter space $\left(\frac{\omega}{\omega_c}, \nu\right)$. The plateau structure for filling factor $\nu = 8$ is marked with a yellow arrow. d) Theoretical predictions of $\theta_F(\frac{\omega}{\omega_c}, \nu)$ showing plateau structure at $\nu = 8$.

**Fig. 2:** a) Cut through the $\theta_F(\nu)$ surface from Fig. 1(c) at $\frac{\omega}{\omega_c} = 0.985$ for $\omega_{THz} = 2.52$ THz (green squares) and $\omega_{THz} = 3.14$ THz (brown dots). Inset shows calculated $\theta_F(\nu)$ for $\frac{\Gamma}{\hbar\omega_c} = 1.1$ (green) and $\frac{\Gamma}{\hbar\omega_c} = 0.7$ (brown). b) Faraday rotation vs. filling factor $\nu$ for various fixed magnetic fields. c.) Comparison between different experiments. DC QHE (blue) is tracked by AC Faraday rotation (red squares; data is scaled to accommodate for Drude background). Faraday rotation vs. filling factor $\nu$ for B=6.62T (purple) shows strong $\nu = 6$ plateau feature. d) Position (bottom, solid symbols) and width (top, open symbols) of the plateaus extracted from $\frac{d\theta_F}{d\nu}$ as a function of $\left(\frac{\omega}{\omega_c}\right)$ (details see Supplementary Information).

**Fig.3:** Faraday rotation as a function of filling factor at $\frac{\omega}{\omega_c} = 0.97$ and at T = 1.7 K (black solid line), T = 2.2 K (red dashed line) and T = 5 K (green dotted line). The inset shows the width of the plateau as a function of temperature. Lines are fits after [5] with exponent $\kappa = 0.5$ (solid), $\kappa = 0.6$ (dashed) and $\kappa = 0.4$ (dotted).